\documentclass[10pt,preprint]{elsarticle}
\usepackage{amsmath}
\usepackage{amssymb}
\usepackage{graphicx}
\usepackage{amsfonts}
\usepackage{graphicx} 
\usepackage[colorlinks=true,urlcolor=blue,citecolor=blue,breaklinks=true]{hyperref}
\usepackage{mathrsfs}
\usepackage{listings}
\usepackage{xparse}

\NewDocumentCommand{\ShowInline}{v}{%
#1%
}

\begin{document}

\begin{frontmatter}

\title{On the number of atomic configurations in hot plasmas}

\author{Jean-Christophe Pain}
\ead{jean-christophe.pain@cea.fr} 
\address{CEA, DAM, DIF, F-91297 Arpajon, France}
\address{Universit\'e Paris-Saclay, CEA, Laboratoire Mati\`ere en Conditions Extr\^emes, F-91680 Bruy\` eres-le-Ch\^atel, France}

\author{Daniel Aberg and Brian G. Wilson}
\address{Lawrence Livermore National Laboratory, P.O. Box 808, L-414, Livermore, California 94551, USA}


\begin{abstract}
We propose approximate and accurate formulas for the number of electron configurations in hot plasmas. Such a quantity is an ingredient of algorithms devoted to the generation of configurations or superconfigurations, which is a pre-requisite of opacity calculations. One of the main formulas involves Bessel functions of fractional order and the procedure for improving the accuracy through a series expansion is explained.
\end{abstract}


\end{frontmatter}

\section{Introduction}\label{sec1}

In a recent paper, Aberg \emph{et al.} proposed a simple approximate formula for the number of atomic configurations \cite{Aberg2024}, \emph{i.e.}, the number of ways to distribute $N$ electrons in an ensemble of $m$ subshells with degeneracies $g_1$, $g_2$, ..., $g_m$. Such a number $\mathscr{N_C}$ is a cornerstone of configuration generators required by hot-plasma opacity calculations \cite{Krief2021,Pain2022,Krief2018,Ovechkin2014}. It can be obtained using recurrence relations, both on the number of subshells and the number of electrons (or holes). However, one often only needs an estimate of such a quantity, not necessarily its exact value. In addition, since this quantity has to be evaluated a large number of times (for example in a superconfiguration generator performing successive gatherings and splittings of supershells), an analytical formula enables one to save computation time. Moreover, it makes it possible to determine (still approximately) the partition of subshells and the number of electrons, for which $\mathscr{N_C}$ does not exceed a ceiling value for instance. 

It turns out that $\mathscr{N_C}$ is the coefficient of $x^N$ in the generating function
\begin{equation}\label{gene}
    f(x)=\prod_{i=1}^m\sum_{j=0}^{g_i}x^j=\prod_{i=1}^m\frac{\left(1-x^{g_i+1}\right)}{\left(1-x\right)}.
\end{equation}
and can expressed as \cite{Pain2020}:
\begin{equation}
    \mathscr{N_C}=\sum_{i=0}^N\binom{i+m-1}{i}
    \sum_{\vec{\alpha}/\sum_{j=1}^m\alpha_j=N-i}\frac{1}{\alpha_1!\alpha_2!...\alpha_m!}
    \prod_{j=1}^m\left(\delta_{\alpha_j,0}-(g_j+1)!\delta_{\alpha_j,g_j+1}\right).
\end{equation}
If we gather the $n_1$ subshells of degeneracy $g_1$, the $n_2$ subshells of degeneracy $g_2$, ..., the $n_s$ subshells of degeneracy $g_s$ (with therefore $n_1+n_2+\cdots n_s=m$), we obtain \cite{Pain2020}:
\begin{equation}\begin{split}
    \mathscr{N_C}=&\sum_{i_1=0}^{n_1}\sum_{i_2=0}^{n_2}\cdots\sum_{i_s=0}^{n_s}
 (-1)^{i_1+i_2+\cdots n_s}\binom{n_1}{i_1}\binom{n_2}{i_2}\cdots\binom{n_s}{i_s} \\
&\times\binom{n_1+\cdots+n_s-1+N-i_1\left(g_1+1\right)-i_2\left(g_2+1\right)-\cdots-i_s\left(g_s+1\right)}{n_1+\cdots+n_s-1}.
\end{split}\end{equation}
An alternate expression can be obtained from sums of powers of roots of a polynomial (see \ref{appA}), but is probably of limited numerical interest. We have also
\begin{equation}
    \mathscr{N_C}=\frac{1}{2\pi i}\oint\frac{1}{z^{N+1}}\prod_{k=1}^m\frac{\left(1-z^{g_k+1}\right)}{\left(1-z\right)}\,\mathrm{d}z.
\end{equation}
A simple approximate expression follows from a unit circle $z=e^{i\theta}$ contour of integration:
\begin{equation}\label{A2}
    \mathscr{N_C}=\frac{1}{2\pi}\int_{-\pi}^{\pi}\prod_{k=1}^m\frac{\left(1-z^{g_k+1}\right)}{z^{Ng_k/G}\left(1-z\right)}\,\mathrm{d}\theta=\frac{1}{2\pi}\int_{-\pi}^{\pi}\prod_{k=1}^m\frac{\left(1-e^{i(g_k+1)\theta}\right)}{e^{iNg_k\theta/G}\left(1-e^{i\theta}\right)}\,\mathrm{d}\theta,
\end{equation}
where we have set $G=\sum_{k=1}^mg_k$. Let us first consider the series expansion of one term in the product of Eq. (\ref{A2}). To second order in $\theta$ we obtain
\begin{align}
    &(1+g_k)+\frac{ig_k(1+g_k)(G-2N)}{2G}\theta\nonumber\\
    &-\frac{g_k(1+g_k)}{12G^2}\left[G^2+2\left(G^ 2-3GN+3N^2\right)g_k\right]\theta^2.
\end{align}
Each term has a common factor of $(1+g_k)$; let us factor it out and define
\begin{equation}
    \alpha=g_k\frac{(G-2N)}{2G}.
\end{equation} 
Then to second order, we have
\begin{equation}
    (1+g_k)\left[1+i\alpha\theta+\left(\beta-\frac{\alpha^2}{2}\right)\theta^2+\cdots\right],
\end{equation} 
where
\begin{equation}
    \beta=-\frac{g_k(2+g_k)}{24}.
\end{equation}
Let us now approximate, as in Ref. \cite{Aberg2024}:
\begin{equation}
    1+i\alpha\theta+\left(\beta-\frac{\alpha^2}{2}\right)\theta^2+\cdots\approx e^{i\alpha\theta+\beta\theta^2},
\end{equation}
and define
\begin{equation}
    G^{(n)}=\sum_{k=1}^mg_k^n.
\end{equation}
Because the integrand with respect to $\theta$ dies quickly, we can replace the limits of integration of $(-\pi,\pi)$ with $(-\infty,\infty)$.  We are left with the analytic Gaussian integral \cite{Hubbard1959,Stratonovich1958}:
\begin{equation}
    \exp \left\{-{\frac {a}{2}}x^{2}\right\}=\frac {1}{\sqrt{2\pi a}}\;\int_{-\infty}^{\infty}\exp \left[-\frac{y^{2}}{2a}-ixy\right]\,dy,
\end{equation}
which yields the formula from Ref. \cite{Aberg2024}, \emph{i.e.},
\begin{equation}\label{f1}
    \mathscr{N}_C\approx\frac{\Lambda}{\frac{\pi}{6}\sqrt{(G^{(2)}+2G)}}\, \exp\left[-\frac{3(G-2N)^2}{2(G^{(2)}+2G)}\right],
\end{equation}
with
\begin{equation}
\Lambda=\prod_{k=1}^m(1+g_k).
\end{equation}
A comparison between the exact number of configurations and the estimate provided by Eq. (\ref{f1}) is displayed in table \ref{tab2} (see the first three columns). We can see that the agreement is rather satisfactory. In the next section, we try to improve the accuracy, while maintaining a certain simplicity.

\section{Improvement of the accuracy}\label{sec2}

\subsection{Third-order term}\label{subsec21}

The third-order term in the series expansion of one term in the product of Eq. (\ref{A2}) reads
\begin{equation}\label{third}
-\frac{i}{24G^3}(1+g_k)(G-2N)g_k^2\left[G^2+\left(G^2-2GN+2N^2\right)g_k\right]\theta^3+o(\theta^4).
\end{equation} 
Since one has the following expansion
\begin{equation}\label{to}
e^{i\alpha\theta+\gamma\theta^2}\approx 1+i\alpha\theta+\left(\beta-\frac{\alpha^2}{2}\right)\theta^2+i\left(\gamma-\frac{\alpha^3}{6}+\alpha\beta\right)\theta^3+\cdots,
\end{equation}
we find, identifying the third-order term of Eq. (\ref{to}) to Eq. (\ref{third}), that $\gamma=0$, which means that there is no contribution from the $\theta^3$ term.

\subsection{Fourth-order term}\label{subsec22}

The fourth-order term in the series expansion of one term in the product of Eq. (\ref{A2}) reads
\begin{align}\label{fourth}
&\frac{(1+g_k)g_k}{720G^4}\left[-G^4+g_k\left(G^4+3G^2\left(3G^2-10GN+10N^2\right)g_k\right.\right.\nonumber\\
&\;\;\;\;\;\;\;\;\;\;\left.\left.+6\left(G^4-5G^3N+10G^2N^2-10GN^3+5N^4\right)g_k^2\right)\right]\theta^4+o(\theta^5),
\end{align} 
and using the following expansion
\begin{equation}\label{to4}
e^{i\alpha\theta+\beta\theta^2+\delta\theta^4}\approx 1+i\alpha\theta+\left(\beta-\frac{\alpha^2}{2}\right)\theta^2+\left(\delta+\frac{\alpha^4}{24}-\frac{\alpha^2\beta}{2}+\frac{\beta^2}{2}\right)\theta^4+\cdots,
\end{equation}
we obtain
\begin{equation}
    \delta=-\frac{1}{2880}\left(4g_k+6g_k^2+4g_k^3+g_k^4\right)
\end{equation}
and thus
\begin{equation}\label{new4}
\mathscr{N}_C=\frac{\Lambda}{2\pi}\int_{-\infty}^{\infty}\exp\left[-\frac{1}{24}(2G+G^{(2)})\theta^2-\frac{1}{2880}\left(4G+6G^{(2)}+4G^{(3)}+G^{(4)}\right)\theta^4\right ]\,\cos\left[\frac{(G-2N)}{G}\theta\right]\,d\theta.
\end{equation}

\begin{table}
    \begin{center}
        \begin{tabular}{cccc}\hline\hline
            $N$ & Exact & Ref. \cite{Aberg2024} (Eq. (\ref{f1})) & This work (Eq. (\ref{sumu}))\\\hline\hline
            5 & 1107 & 2128 & 1304\\
            10 & 20469 & 19903 & 20789\\
            15 & 95580 & 90496 & 94412\\
            20 & 196425 & 200052 & 196995\\
            25 & 209270 & 215012 & 210297\\
            30 & 116883 & 112354 & 115756\\
            35 & 30265 & 28544.5 & 30289\\
            40 & 2349 & 3526 & 2699\\\hline\hline  
        \end{tabular}
        \caption{Exact number of configurations and approximate formulas (\ref{f1}) (from Ref. \cite{Aberg2024}) and (\ref{sumu}) (this work) for different values of the number of electrons $N$ in the case of the 9 first non-relativistic subshells (1s to 4d).}\label{tab2}
    \end{center}
\end{table}

\begin{figure*}[ht!]
\centering
\includegraphics[width=10cm]{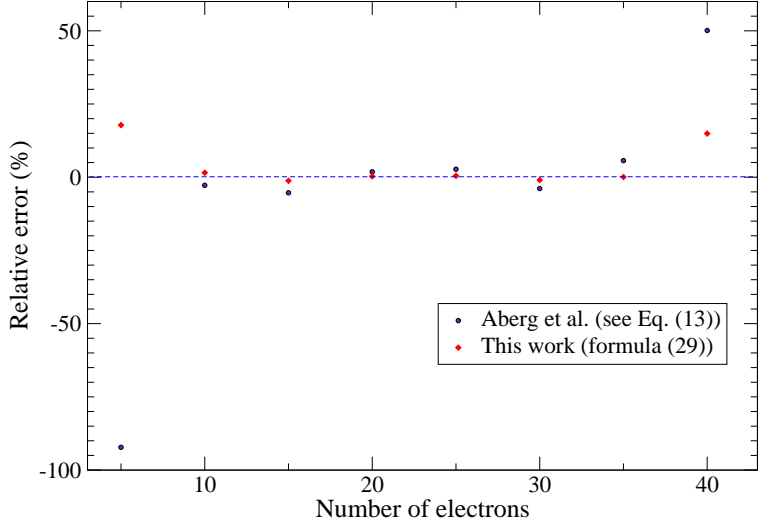}
\caption{\label{fluctu} Relative error ($\mathscr{N}_C$[approx]-$\mathscr{N}_C$[exact])/$\mathscr{N}_C$[exact], where $\mathscr{N}_C$[approx] is calculated using the expression of Aberg \emph{et al.} (see Eq. (\ref{f1}) and using formula (\ref{sumu}) (or any evaluation of formula (\ref{new4})) as a function of the number of electrons in the supershell (1s to 4d).}
\end{figure*}

Expression (\ref{new4}) can be easily implemented numerically. Rewriting the preceding equation with the variable $\alpha$, $\beta$ and $\delta$, one has
\begin{equation}
\mathscr{N}_C=\frac{\Lambda}{2\pi}\int_{-\infty}^{\infty}\exp\left(\beta\theta^2+\delta\theta^4\right )\,\cos\left(\alpha\theta\right)\,d\theta.
\end{equation}
Using the series expansion
\begin{equation}
\cos(\alpha\theta)=\sum_{p=0}^{\infty}\frac{(-1)^{p}}{(2p)!}\,\alpha^{2p}\,\theta^{2p},
\end{equation}
one obtains
\begin{equation}
\mathscr{N}_C=\frac{\Lambda}{2\pi}\sum_{p=0}^{\infty}\frac{(-1)^{p}}{(2p)!}\alpha^{2p}\int_{-\infty}^{\infty}\exp\left(\beta\theta^2+\delta\theta^4\right)\,\theta^{2p}\,d\theta.
\end{equation}
One has actually
\begin{equation}
\int_{-\infty}^{\infty}\exp\left(\beta\theta^2+\delta\theta^4\right)\,\theta^{2p}\,d\theta=\frac{\Gamma(p+1/2)}{2^{p+1/2}(-\delta)^{(2p+1)/4}}U\left(\frac{2p+1}{4},\frac{1}{2};-\frac{\beta^2}{4\delta}\right),
\end{equation}
where
\begin{equation}
    U(a,b;z)=\frac{1}{\Gamma(a)}\int_0^{\infty}e^{-zt}t^{a-1}(1+t)^{b-a-1}\,dt
\end{equation}
is the Tricomi confluent hypergeometric function, also referred to as the confluent hypergeometric function of the second kind. $\Gamma$ is the usual Gamma function. One has in particular
\begin{equation}
U(a,b;z)=\frac{\pi}{\sin(\pi b)}\left[\frac{~_1F_1(a,b;z)}{\Gamma(1+a-b)\Gamma(b)}-z^{1-b}\,\frac{~_1F_1(a+1-b,2-b;z)}{\Gamma(a)\Gamma(2-b)}\right]
\end{equation}
or also
\begin{equation}
U(a,b;z)=\frac{\Gamma(1-b)}{\Gamma(a+1-b)}~_1F_1(a,b;z)+\frac{\Gamma(b-1)}{\Gamma(a)}z^{1-b}~_1F_1(a+1-b,2-b;z)
\end{equation}
where $_1F_1$ is Kummer's hypergeometric function of the confluent hypergeometric function of the first kind (also denoted $M(a,b;z)$). The following {\sc Python} code can be used to convert one to another: from \verb!scipy.special!, import \verb!gamma!, \verb!hyp1f1! (via \verb!scipy.special.hyp1f1!):

\begin{verbatim}
def U(a, b, z):
    # First term of the equation
    term1 = gamma(1 - b) / gamma(a + 1 - b) * hyp1f1(a, b, z)

    # Second term of the equation
    term2 = gamma(b - 1) / gamma(a) * z**(1 - b) * hyp1f1(a + 1 - b, 2 - b, z)

    # Combine the terms
    result = term1 + term2
    return result

print(U(5/4,1/2,1/4))
\end{verbatim}

It is worth mentioning that an algorithm in MATLAB for the computation of the Tricomi function was proposed in Ref. \cite{Gil2023}. Such an idea (expanding the cosine function) is also of interest for the computation of the Voigt profile (see the work of Zaghloul \cite{Zaghloul2007}), although the integral is different in the latter case, since one has
\begin{equation}
H(a,v)=\frac{a}{\pi}\int_{-\infty}^{\infty}\frac{e^{-y^2}}{(v-y)^2+a^2}dy=\frac{1}{\sqrt{\pi}}\int_0^{\infty}\exp\left(-ax-\frac{x^2}{4}\right)\cos(vx)dx.
\end{equation}
Finally, one gets
\begin{equation}\label{sumu}
\mathscr{N}_C=\frac{\Lambda}{2\pi}\sum_{p=0}^{p_{\mathrm{max}}}\frac{(-1)^{p}}{(2p)!}\frac{\alpha^{2p}\,\Gamma(p+1/2)}{2^{p+1/2}(-\delta)^{(2p+1)/4}}\,U\left(\frac{2p+1}{4},\frac{1}{2};-\frac{\beta^2}{4\delta}\right).
\end{equation}

\begin{table}
    \centering
    \begin{tabular}{cc}\hline\hline 
    $p_{\mathrm{max}}$ & $\mathscr{N}_C$ \\\hline\hline 
    5 & 7278\\
    6 & 24180\\
    7 & 20071\\
    8 & 20919\\
    9 & 20768\\
    10 & 20792\\
    11 & 20788.5\\
    12 & 20789\\
    13 & 20789\\\hline\hline    
    \end{tabular}
    \caption{Value of the number of configurations for $N$=10 for different values of the maximum index of summation entering Eq. (\ref{sumu}).}\label{expan}
\end{table}

\begin{table}
    \centering
    \begin{tabular}{cc}\hline\hline 
    $p_{\mathrm{max}}$ & $\mathscr{N}_C$ \\\hline\hline 
    0 & 221545\\
    1 & 195608\\
    2 & 197044\\
    3 & 196993\\
    4 & 196995\\
    5 & 196995\\\hline\hline    
    \end{tabular}
    \caption{Value of the number of configurations for $N$=20 for different values of the maximum index of summation entering Eq. (\ref{sumu}).}\label{expan2}
\end{table}

The convergence is obtained very quickly, especially close to half-filled supershell (see tables \ref{expan} and \ref{expan2}). Each term in the summation can be expressed as a combination of Bessel functions of fractional order. For instance, for the first two terms one has
\begin{equation}
\mathscr{N}_C\approx\frac{\Lambda\,\alpha}{2\sqrt{2\pi}\,(-\delta)^{1/4}}\,U\left(\frac{1}{4},\frac{1}{2};-\frac{\beta^2}{4\delta}\right)-\frac{\Lambda\,\alpha^2}{16\sqrt{2\pi}\,(-\delta)^{3/4}}\,U\left(\frac{3}{4},\frac{1}{2};-\frac{\beta^2}{4\delta}\right)+\cdots
\end{equation}
and since
\begin{equation}
U\left(\frac{1}{4},\frac{1}{2};x\right)=\frac{e^{x/2}\,x^{1/4}}{\sqrt{\pi}}\,K_{1/4}\left(\frac{x}{2}\right)
\end{equation}
as well as
\begin{equation}
U\left(\frac{3}{4},\frac{1}{2};x\right)=\frac{2\,e^ {x/2}}{\sqrt{\pi}\,x^{1/4}}\left[-(1+x)\,K_{1/4}\left(\frac{x}{2}\right)+x\,K_{5/4}\left(\frac{x}{2}\right)\right]
\end{equation}
where $K_{1/4}$ and $K_{5/4}$ are Bessel functions of fractional order, one gets
\begin{equation}
\mathscr{N}_C\approx\frac{\Lambda}{32\pi\delta}\,e^{-\beta^2/(8\delta)}\left[\left(\frac{\alpha^2(4\delta-\beta^2)}{\sqrt{\beta\delta}}-8\sqrt{\beta\delta}\right)\,K_{1/4}\left(-\frac{\beta^2}{8\delta}\right)+\frac{\alpha^2\beta^2}{\sqrt{\beta\delta}}\,K_{5/4}\left(-\frac{\beta^2}{8\delta}\right)\right]+\cdots
\end{equation}
The same patterns remains true for the next terms; for instance:
\begin{equation}
U\left(\frac{5}{4},\frac{1}{2};x\right)=\frac{4\,e^ {x/2}}{3\sqrt{\pi}\,x^{3/4}}\left[-(3+2x(3+x))\,K_{3/4}\left(\frac{x}{2}\right)+x(1+2x)\,K_{7/4}\left(\frac{x}{2}\right)\right].
\end{equation}
and
\begin{equation}
U\left(\frac{7}{4},\frac{1}{2};x\right)=\frac{8\,e^ {x/2}}{15\sqrt{\pi}\,x^{5/4}}\left[(15+x(45+x(23+4x)))\,K_{5/4}\left(\frac{x}{2}\right)-x(3+x(9+4x))\,K_{9/4}\left(\frac{x}{2}\right)\right].
\end{equation}

Figure \ref{fluctu} represents the relative error ($\mathscr{N}_C$[approx]-$\mathscr{N}_C$[exact])/$\mathscr{N}_C$[exact], where $\mathscr{N}_C$[approx] is calculated using the formula of Aberg \emph{et al.} (see Eq. (\ref{f1})) and using formula (\ref{sumu}) (or any evaluation of Eq. (\ref{new4})), for different values of the number of electrons, still in the supershell (1s to 4d). We can see that the accuracy is always improved with the new formula. This is confirm by the numerical values in the last column of table \ref{tab2}.

\subsection{Generalization: towards a more accurate approximation}\label{subsec23}

Higher orders can be obtained simply by noticing that the odd powers of $\theta$ do not contribute in the exponential, and that the logarithm of the term multiplying $\theta^{2p}$, where $p$ is an integer, reads
\begin{equation}
-\prod_{k=0}^{p+1}\frac{k!}{(2k)!}\sum_{j=1}^{2p}\binom{2p}{j}G^{(j)},
\end{equation}
where we recognize coefficients of the Pascal triangle without 1 (the $j=0$ case). Thus, the approximate formula for the number of configurations up to order $2r$ reads
\begin{equation}
\mathscr{N}_C=\frac{\Gamma}{2\pi}\sum_{p=0}^{\infty}\frac{(-1)^{p}}{(2p)!}\int_{-\infty}^{\infty}\exp\left[-\sum_{q=1}^r\left(\prod_{k=0}^{q+1}\frac{k!}{(2k)!}\right)\left(\sum_{j=1}^{2q}\binom{2q}{j}G^{(j)}\right)\theta^{2q}\right]\,\theta^{2p}\,d\theta
\end{equation}
and we can write
\begin{equation}\label{coefrac}
\prod_{k=0}^p\frac{k!}{(2k)!}=\frac{2^{\frac{1}{24}-p(p+1)}\,e^{1/8}\,\pi^{\frac{2p+1}{4}}}{A^{3/2}\,\mathrm{G}(p+3/2)},
\end{equation}
where $A\approx 1.28243$ is the Glaisher-Kinkelin constant \cite{Glaisher1878,Finch2003} equal to
\begin{equation}
\ln(A)=\frac{1}{12}-\zeta'(-1),
\end{equation}
$\mathrm{G}$ being the so-called Barnes $\mathrm{G}$-function \cite{Barnes1900,Adamchik2003} and $\zeta'$ the derivative of the Riemann zeta function $\zeta$. The inverse of the coefficient in the left-hand side of Eq. (\ref{coefrac}) is equal to 2 for $p=1$, to 24 for $p=2$, to 2880 for $p=3$, to 4838400 for $p=4$ and to 146,313,216,000 for $p=5$.

\section{Conclusion}\label{sec3}

In this note, we derived approximate analytical formulas for the number of atomic configurations in hot plasmas. Such a quantity is important for algorithms devoted to the generation of a list of detailed ionic configurations of bound states for all ion stages in the plasma, which is a cornerstone of opacity calculations. Although the number of atomic configurations can be obtained from efficient recurrence relations, having a fast formula not only saves computational ressources, as this quantity has to be evaluated a large number of times, but also makes it possible to determine the parameters (number and degeneracy of subshells, number of electrons, \emph{etc.}) for which the number of configurations does not exceed a limit value.

\appendix

\section{An alternate expression for the number of configuration}\label{appA}

The well-known identity (where we assume that $n$ is an even integer)
\begin{equation}
\sum_{k=0}^{n/2}(-1)^k\frac{n}{n-k}\binom{n-k}{k}(x+y)^{n-2k}(xy)^k=x^n+y^n
\end{equation}
is a special case of an older formula for sums of powers of roots of a polynomial which was first found by Girard and later given by Waring \cite{Gould1999}. Its dual form is
\begin{equation}
\sum_{k=0}^{n/2}(-1)^k\binom{n-k}{k}(x+y)^{n-2k}(xy)^k=\frac{x^{n+1}-y^{n+1}}{x-y}.
\end{equation}
Setting $y=1$ yields
\begin{equation}
\frac{1-x^{n+1}}{1-x}=\sum_{k=0}^{n/2}(-1)^k\binom{n-k}{k}(x+1)^{n-2k}\,x^k
\end{equation}
and thus the generating function $f(x)$ (see Eq. (\ref{gene})) becomes
\begin{equation}
f(x)=\prod_{i=1}^m\left(\sum_{k_i=0}^{g_i/2}(-1)^{k_i}\binom{g_i-k_i}{k_i}(x+1)^{g_i-2k_i}\,x^{k_i}\right)
\end{equation}
or equivalently
\begin{equation}
f(x)=\sum_{k_1=0}^{g_1/2}\sum_{p_1=0}^{g_1-2k_1}\sum_{k_2=0}^{g_2/2}\sum_{p_2=0}^{g_2-2k_2}\cdots \sum_{k_m=0}^{g_m/2}\sum_{p_m=0}^{g_m-2k_m}(-1)^{\sum_{j=1}^mk_j}\prod_{i=1}^m\binom{g_i-k_i}{k_i}\binom{g_i-2k_i}{p_i}\,x^{p_i+k_i},
\end{equation}
and thus the number of configurations reads
\begin{equation}
\mathscr{N}_c=\underbrace{\sum_{k_1=0}^{g_1/2}\sum_{p_1=0}^{g_1-2k_1}\sum_{k_2=0}^{g_2/2}\sum_{p_2=0}^{g_2-2k_2}\cdots \sum_{k_m=0}^{g_m/2}\sum_{p_m=0}^{g_m-2k_m}}_{\sum_{j=1}^m(k_j+p_j)=N}(-1)^{\sum_{j=1}^mk_j}\prod_{i=1}^m\binom{g_i-k_i}{k_i}\binom{g_i-2k_i}{p_i}.
\end{equation}

\section*{Acknowledgements}

J.-C. Pain is indebted to X. Blanc for helpful discussions. 
This work was performed under the auspices of the U.S. Department of Energy by Lawrence Livermore National Laboratory under Contract DE-AC52-07NA27344.

\end{document}